\documentclass[twocolumn,prd,preprintnumbers,amsmath,amssymb,superscriptaddress]{revtex4}
\usepackage{amsmath}
\usepackage{graphicx}
\usepackage{color}
\newcommand{\comment}[1]{}

\newcommand{\Hess}{{\cal H}}

\newcommand{\nn}{\nonumber\\}

\usepackage[normalem]{ulem}  

\renewcommand\sout{\bgroup \color{red} \ULdepth=-.5ex \ULset}
\newcommand\soutb{\bgroup \color{blue} \ULdepth=-.5ex \ULset}

\begin{document}

\preprint{KUNS-2433 / YITP-13-10}

\title{Entropy production in classical Yang-Mills theory from Glasma initial conditions}

\author{Hideaki Iida}
\affiliation{Department of Physics, Kyoto University, Kyoto 606-8502, Japan}
\author{Teiji Kunihiro}
\affiliation{Department of Physics, Kyoto University, Kyoto 606-8502, Japan}
\author{Berndt M\"uller}
\affiliation{Department of Physics \& CTMS, Duke University, Durham, NC 27708, USA}
\author{Akira Ohnishi}
\affiliation{Yukawa Institute for Theoretical Physics, Kyoto University,
Kyoto 606-8502, Japan}
\author{Andreas Sch\"afer}
\affiliation{Institut f\"ur Theoretische Physik, Universit\"at Regensburg,
D-93040 Regensburg, Germany}
\affiliation{Yukawa Institute for Theoretical Physics, Kyoto University,
Kyoto 606-8502, Japan}
\author{Toru T. Takahashi}
\affiliation{Gumma National College of Technology, Gumma 371-8530, Japan}

\date{\today}


\begin{abstract}

We study the thermalization process in classical Yang-Mills (CYM) field theory 
starting from noisy glasma-like initial conditions
by investigating the initial-value sensitivity of trajectories. 
Kunihiro {\it et al.} \cite{Kunihiro:2008gv} linked entropy generation  
to the Kolmogorov-Sina\"i entropy, which
gives the entropy production rate in classical chaotic systems,
calculated numerically for CYM fields
starting from purely random initial field configurations. 
In contrast, we here study glasma-like initial conditions.
For small random fluctuations we obtain qualitatively similar results 
while no entropy increase is observed when such fluctuations are absent. 
We analyze the intermediate time Lyapunov spectrum for several time windows 
and calculate the Kolmogorov-Sina\"i entropy.
We find a large number of positive Lyapunov exponents
at the early stages of time evolution. Also for later times their
number is a sizeable fraction of the total number of degrees of freedom.
The spectrum of positive Lyapunov exponents at first changes
rapidly, but then stabilizes, indicating that the dynamics of the
gauge fields approaches a steady state.
Thus we conclude that also for glasma-like initial conditions 
a significant amount of entropy is produced by  
classical gluon field dynamics.
\end{abstract}


\maketitle
\section{Introduction}
\label{sec1}

\subsection{Motivation}

Elucidating the mechanism of entropy production and early thermalization  
leading to  quark-gluon plasma (QGP) formation
is one of the fundamental problems posed by 
the phenomenological analyses of the experimental data
obtained from the Relativistic Heavy-Ion Collider (RHIC)
at Brookhaven National Laboratory 
and the Large Hadron Collider (LHC) at  CERN.
As hydrodynamics is only applicable after local thermalization
has occurred, massive entropy production is a basic
ingredient for any detailed understanding of the time-evolution of 
heavy-ion collisions.
Analyses based on ideal relativistic hydrodynamic equations 
suggest that thermalization should be achieved in an early stage 
in order to explain the RHIC data.
The  thermalization time is estimated to be $\tau_0\simeq 1 ~\mathrm{fm}/c$ 
\cite{heinz,HK2001}, which is significantly 
shorter than the equilibration time obtained in perturbative QCD \cite{Baier:2000sb}. 
If $\tau_0$ were much longer, one would have to expect 
that significant artifacts caused by incomplete thermalization  
could affect many interpretations of the  experimental data. 

The ``early thermalization'' of the created matter 
should be explained on the basis of the underlying dynamics
of the time-evolution of the matter created in the nuclear collision. 
This remains a major challenge of relativistic heavy ion physics.
Apart from the time-evolution dynamics, peculiar fluctuations in the
{\em initial conditions} are also recognized to be an important ingredient
for understanding the hydrodynamic evolution of the quark gluon matter,
in particular, the elliptic and triangular flows
\cite{Alver:2010dn,Petersen:2010cw,Dumitru:2012yr,Dusling:2012yd}.

\subsection{Classical Yang-Mills Fields}

Classical Yang-Mills (CYM) field theory is 
a good starting point for describing the initial stage dynamics 
of relativistic heavy-ion collisions.
At high energies,  wee partons with small Bjorken $x$ are excited,
and the number of gluons in the small $x$ region is so high
that they may be treated as a coherent or classical field
\cite{CGCreview,MV1994}.
A classical solution of gluon fields in this regime is known: 
A nucleus at high energy can be regarded as a collection
of large-$x$ particles, the valence partons,
which act as color sources that give rise to 
strong, transversely polarized chromo-magnetic and -electric fields. 
This configuration of the gluon fields is referred to as the color glass condensate (CGC).
When two nuclei collide, their gluon fields interact and produce
new, longitudinally polarized fields that extend between the
color sources in the direction of the beam axis. This {\em glasma}
field configuration is used as a standard 
initial condition for the time evolution of the matter created in 
relativistic heavy-ion collisions.

It is generally expected that 
some field instabilities could cause early thermalization; 
such instabilities include the Weibel instability \cite{W1959,Mrowczynski:1993qm} 
and Nielsen-Olesen instability \cite{Nielsen:1978rm,FI2008,FII2009}. 
In both cases, the instability gives rise to an exponential growth
of the amplitude of unstable modes having high energy,
which are expected to cascade
into a large number of modes or particles with small energies
due to the nonlinear coupling of fields,
until the fields are thermalized.

There are two facets in thermalization of CYM.
The first aspect is the isotropization of the energy momentum tensor.
The approximate isotropization is the minimal condition to apply hydrodynamics,
and has been discussed in the literature~\cite{Dumitru:2006pz,Lappi:2006fp}.
The second is the equilibration of the spectrum toward the Bose-Einstein
and the Fermi-Dirac distribution 
of gluons and quarks, respectively.
While the spectrum equilibration provides for a more rigorous definition
of thermalization, it has not been discussed seriously so far in CYM.
One reason for this may be that classical field theories do not generally 
exhibit the correct equilibrium spectrum by themselves.
When equipartition of energy is realized in classical equilibrium,
high momentum components are favored compared
with the quantum equilibrium, i.e. the Bose distribution.
In the thermalization scenario described in the previous paragraph,
we expect the produced particles to show quantum equilibrium spectra.
Thus the very thermalization mechanism is attributed to this conversion 
process but it is a difficult and unsolved problem 
to describe 
non-equilibrium and non-uniform situations
in terms of nonlinear classical or quantum dynamics.

\subsection{Entropy Production}

A central task
to reach an 
understanding of early thermalization is
to identify the mechanism of entropy production, which has been scarcely 
touched upon so far in the literature.  It is to be noted that
the high occupation probability of the unstable modes makes 
possible a
quasi-classical treatment of the thermalization process in 
the initial stage.
Furthermore, Kunihiro, M\"uller, Ohnishi, and Sch\"afer \cite{Kunihiro:2008gv} 
noticed that the use of 
the Husimi-Wehrl entropy $S_{\mathrm{HW}}$, 
the Wehrl entropy~\cite{Wehrl:1978zz} defined in terms of
a smeared Wigner function (the Husimi function)~\cite{Husimi:1940}, 
is meaningful in the quasi-classical regime 
and showed that $S_{\mathrm{HW}}$ grows
at the rate of the Kolmogorov-Sina\"i (KS) entropy $S_\mathrm{KS} $
in the long-time limit.
The fundamental origin of entropy growth induced by the Husimi transformation 
is that even without an actual measurement the quantum mechanical uncertainty 
principle implies a minimum of coarse-graining.
In other words: Information which cannot be measured in accordance with the 
uncertainty principle is de facto 
lost and information loss is equivalent to entropy growth.  

Here the KS entropy is defined as a sum of the positive Lyapunov exponents
$\lambda_i$,
\begin{align}
S_{\mathrm{KS}} = \sum_{i, \lambda_i > 0} \lambda_i\ .
\end{align}
The size of a Lyapunov exponent $\lambda_i$ is an indicator
for the initial-value sensitivity of trajectories
defined through the equation, 
$|\delta X_i(t)| \simeq |\delta X_i(t_0)|\, \exp[\lambda_i (t-t_0)]$,
where $\delta X_i$ represents a small difference of phase-space variables  
at the initial time $t=t_0$.
A positive Lyapunov exponent means that the distance between
the two phase space points grows exponentially in time.
Thus the entropy production in classical dynamics
is closely related to the chaoticity of the system.
It is therefore quite pertinent to 
investigate possible entropy production
of the classical field itself to study the thermalization mechanism through
its chaotic behavior in the glasma stage.
 
In fact, the study of the chaotic properties of the classical evolution of 
Yang-Mills fields has  a long history
\cite{Biro:1994bi}, initiated by the observation of chaotic behavior 
in the infrared limit of Yang-Mills theories \cite{Matinyan:1986nw},
which was confirmed later
for the compact lattice formalism
of classical Yang-Mills
theory \cite{Muller:1992iw}.  Some properties of Lyapunov exponent 
and KS entropy in compact lattice gauge theories
are discussed in \cite{Biro:1994sh} and \cite{Gong:1993xu}, respectively.

In Ref. \cite{KMOSTY2010},  the authors 
analyzed the exponential growth of the distance between  two trajectories 
for classical Yang-Mills evolution, 
starting from  adjacent {\em generic} random initial gauge fields 
in the noncompact $(A,E)$ scheme.  They calculated the KS entropy and
found that the KS entropy is positive and finite even after a long time.
The equilibration time scale $\tau_\mathrm{eq}$ was 
estimated to be around 2 fm$/c$ for $T = 350~\mathrm{MeV}$,
though with rather substantial systematic uncertainties. 
This result obtained by starting from a generic random initial fields 
shows that a significant amount of entropy is produced by 
the dynamical complexity inherent in the CYM equations,
suggesting that thermalization in heavy-ion collisions 
can be at least partly be achieved 
in the classical regime before 
particle production comes into play as a quantum process.

In Ref.~\cite{KMOSTY2010}, 
it was also observed  
that there are three distinct time regimes,
namely a kinetic stage for short sampling times, an intermediate- and 
a long-time regime:\,
i)~ The short sampling time
   is characterized by the {\em local Lyapunov exponents} (LLE).\,
ii)~The evolution of the distance on (long) mixing time scales is described by the
 usual Lyapunov exponents,  which we  refer to as {\em global Lyapunov exponents} (GLE). 
It is to be noted that GLEs are equivalent to the original Lyapunov exponents.
iii)~In the intermediate-time period, the nonlinear coupling 
between different field modes is significant
but the energy remains localized among the primary unstable modes. 
The Lyapunov exponents characterizing the exponential growth of the 
separation of trajectories in this intermediate time period are
called {\em intermediate Lyapunov exponents} (ILEs). 
We emphasize that the ILEs are the most relevant Lyapunov exponents for 
the thermalization of the glasma, because they characterize
the time-evolution of the strongly excited Yang-Mills fields
in the early stage
when the field configuration is still far away from equilibrium
and a quasi-classical description of the dynamics of the Yang-Mills field
is appropriate.
 
\subsection{Scope of this Work}

In this paper, we extend the analysis done in \cite{KMOSTY2010}
and explore the thermalization process starting from more 
realistic initial conditions than those adopted in Ref.~\cite{KMOSTY2010},
namely  glasma-like initial conditions. 
We analyze
the time-evolution of the distance between two trajectories starting from 
two adjacent points in phase space and also extract the spectrum of 
Lyapunov exponents from the time evolution of these fields. 
As was mentioned before, we can determine whether and how early entropy production is
achieved by examining the number and magnitude of the positive Lyapunov exponents. 
In this study, we use two glasma-like initial conditions.
One is called ``modulated initial condition'',
where the initial color-magnetic fields $B_i$ are spatially modulated
along the $z$ and $x$-axes. 
We call the other ``constant-$A$ initial condition''. Here both the 
gauge potentials $A_i$ and the chromomagnetic fields $B_i$ are constant,
but non-commuting.  
It turns out that the two initial conditions give similar 
results for entropy production with some minor differences in the 
time evolution once the glasma-like initial conditions are taken.

The paper is organized as follows. 
In Sec.~\ref{sec2}, we introduce
the basic ingredients of our simulations including 
the initial conditions.
In Sec.~\ref{sec3}, we show the numerical results
for calculations starting from the modulated initial condition.
From the time evolution of the distance between two trajectories
which are very close to each other initially 
we determine the Lyapunov exponents. 
We also show numerical results for simulations using   
the constant-$A$ initial condition in Sec.~\ref{sec3}. 
Section ~\ref{sec5} is devoted to a summary and concluding remarks. 

\section{Formulation of the Problem}
\label{sec2}

In this section, we first introduce the equations of motion and
the formulae needed for the analysis of chaotic behavior
and the entropy production.
(See Ref.~\cite{KMOSTY2010} for details.) 
Then we describe the glasma-like initial conditions
with fluctuations as well as the parameters chosen in our simulations.

\subsection{Classical Yang-Mills equation}

In pure Yang-Mills theory in temporal gauge $A^a_0=0$,
the Hamiltonian in the noncompact $(A,\, E)$ scheme 
takes the following form on a cubic spatial  lattice
\begin{align}
H=&\frac{1}{2}\sum_{x,a,i}E_i^a(x)^2 + \frac{1}{4}\sum_{x,a,i,j}F_{ij}^a(x)^2 
\ ,\label{eq1} \\
F_{ij}^a(x) &= \partial_i A_j^a (x)-\partial_j A_i^a (x)
+ \sum_{b,c} f^{abc}A^b_i(x)A^c_j(x)
\ , \label{eq2}
\end{align}
where $\partial_i$ is the central difference operator in the $i$-direction,
$\partial_i A(x) \equiv \{ A(x+\hat{i})-A (x-\hat{i}) \}/2$, and 
$f^{abc}$ are structure constants. In this study, we deal with SU(2) gauge theory, 
i.e., $f^{abc}=\epsilon^{abc}$, where $\epsilon^{abc}$ is 
the Levi-Civita tensor defined with $\epsilon^{123}=1$.   
Note that all quantities in the equations are dimensionless, i.e. scaled with appropriate 
powers of the lattice constant. 
  
From Eqs.~\eqref{eq1} and \eqref{eq2}, we get the classical 
equations of motion (EOM)
for $(A_i^a(x),\,E_i^a(x))$ 
\begin{align}
\dot{A}_i^a(x) &= E_i^a(x) \ ,\\
\dot{E}_i^a(x) &= \sum_{j} \partial_j F_{ji}^a(x)+ \sum_{b,c,j} f^{abc}A^b_j(x) F_{ji}^c(x)
\ ,
\end{align}
which we solve with glasma-like initial conditions to be specified later. 
A fourth-order Runge-Kutta method is adopted to solve these EOM. 
It should be noted that we chose the initial condition 
such that it satisfies Gauss' law 
$(\sum_i D_i E^i(x))^a=\sum_{i} \partial_i E^{ia}(x)+ \sum_{i, c, b}f^{acb}A_i^c E^{ib}(x)=0$ 
and check its validity carefully at every time slice. 
Energy conservation is also checked for each time step. 

\subsection{Distance between two trajectories}

A natural indicator for the chaotic behavior of the system is the ``distance'' between two trajectories
which start from slightly different initial points in phase space. 
With $(A^a_i(t,\, \vec {r}),\,E^a_i(t,\, \vec{r}))$ being the trajectory
starting from an initial point $(A^a_i(0,\, \vec {r}),\,E^a_i(0,\, \vec{r}))$ in phase space,
we consider a second trajectory $(A^{\prime a}_i(t,\, \vec {r}),\,E^{\prime a}_i(t,\, \vec{r}))$
starting from an adjacent point.
Then we define  the distance $D_{EE}$ ($D_{FF}$) between the electric fields (the 
field strengths) by
\begin{align}
D_{EE}=\sqrt{\sum_x\left\{\sum_{a,i} E_i^a(x)^2-\sum_{a,i}E_i^{\prime a}(x)^2\right\}^2},\\
D_{FF}=\sqrt{\sum_x\left\{\sum_{a,i,j} F_{ij}^a(x)^2-\sum_{a,i,j}F_{ij}^{\prime a}(x)^2\right\}^2},
\end{align}
respectively.  
Here 
$F^{\prime a}_{ij}(x)$ is
the  field strength tensor
evolved from the initial point 
$(A^{\prime a}_i(t=0, \vec{r}),\, E^{\prime a}_i(t=0,\,\vec{r}))$. 
These distances are gauge invariant under the residual gauge transformation of 
the EOM, namely,  
$E \rightarrow \Omega(\vec x) E$ 
and 
$F \rightarrow \Omega(\vec x) F$,
where $\Omega(\vec x)$ is a function of 
gauge transformations in the adjoint representation which are independent of time. 

The Lyapunov exponents are extracted from the time-dependence of these distances, as
is described in the next subsection.

\subsection{Lyapunov exponents} 

For the two trajectories 
$(A_i^a(t,\, \vec{x}),\, E_i^a(t,\, \vec{x}))$, 
and $({A_i^a}'(t,\, \vec{x}),\, {E_i^a}'(t,\, \vec{x}))$ 
the tangent vector 
\begin{align}
\delta X(t) = (\delta A_i^a(t,\, \vec{x}), \delta E_i^a(t,\, \vec{x}))^T
\end{align}
satisfies the following EOM\cite{KMOSTY2010},
\begin{align}
\delta \dot{X}(t) =
\begin{pmatrix}
{\bf 0} & {\bf 1} \\
-H_{AA}(t) & {\bf 0} \\
\end{pmatrix}
\delta X(t)
\equiv \Hess(t) \delta X(t) \ ,
\label{eq5}
\end{align}
where we have introduced a matrix defined by
\begin{align}
(H_{AA}(t))_{iax,jby}&=\delta^2H/\delta A_i^a(\vec x,t) \delta A_j^b(\vec y,t).
\end{align}

We call the matrix ${\cal H}$ Hessian, and 
the eigenvalues of $\Hess$ {\em for each time slice} are referred
to as the local Lyapunov exponents (LLEs)\cite{KMOSTY2010}:
As it stands,
the LLE plays the role of a ``temporally local'' Lyapunov exponent,
which specifies the departure rate of two trajectories in a short time period. 

For a system where stable and unstable modes couple with each other
as in the present case,
an LLE does not generally agree with the Lyapunov exponent in a long time period.
As an adequate way to characterize the exponential growth of the fluctuation,
the authors in Ref.~\cite{KMOSTY2010}  introduced
another kind of Lyapunov exponent $\lambda^{\rm ILE}$ 
called the intermediate Lyapunov exponent (ILE),
which is an ``averaged Lyapunov exponent'' for an intermediate time period $\Delta t$;
i.e., a time period which is sufficiently small compared to the thermalization time
but large enough to sample a significant fraction of phase space.

The explicit definition of a $\lambda^{\rm ILE}$ goes as follows:
We first note that
Eq.~\eqref{eq5} is solved for any time period $\Delta t$ by
\begin{align}
\delta X(t+\Delta t)
=\,&U(t,t+\Delta t) \delta X(t)
\ ,\\
U(t,t+\Delta t)
=\,& {\cal T} \left[\exp\left( \int_t^{t+\Delta t} \Hess(t+t') dt' \right)	\right]
\ ,
\end{align}
with ${\cal T}$ denoting the time ordered product.
The following Trotter formula
for the time-evolution operator $U$
is found convenient for a numerical evaluation;
\begin{align}
U(t,t+\Delta t)=\,& {\cal T} \prod_{k=1,N}
U(t_{k-1},t_k)
\nn
\simeq\,& {\cal T} \prod_{k=1,N}
\left[1+ \Hess(t_{k-1}) \delta t
\right]
\ ,
\label{Eq:Trotter2}
\end{align}
where $\delta t \equiv \Delta t/N$.
Diagonalizing the matrix $U$, we reach the definition of the ILEs 
\begin{align}
U_D(t,t+\Delta t) =
\mathrm{diag}
(e^{\lambda^\mathrm{ILE}_1 \Delta t}, e^{\lambda^\mathrm{ILE}_2 \Delta t}, \ldots).
\end{align}
In this study, $\lambda^{\rm ILE}$'s are calculated by setting $t=0$,
\begin{align}
U_D(0,\Delta t) =
\mathrm{diag}
(e^{\lambda^\mathrm{ILE}_1 \Delta t}, e^{\lambda^\mathrm{ILE}_2 \Delta t}, \ldots).
\end{align}
As the thus defined ILEs depend on the time period $\Delta t$,
we examine the $\Delta t$ dependence of the ILE spectrum 
in our later discussion.

Two comments are in order, here: 
A Lyapunov exponent can be (real) positive, negative, zero or purely 
imaginary.
Liouville's theorem tells us that
the determinant of the time evolution matrix $U$ is unity,
implying that the sum of all positive and negative ILEs is zero.
The KS entropy is given as a sum of positive Lyapunov exponents. 
The second comment concerns gauge invariance of the Lyapunov exponents. 
The discussion in the paper is centered on the time evolution of 
KS entropy which is the sum of the 
positive Lyapunov exponents in CYM. 
However, if the modes related to LLE and ILE are gauge non-invariant, 
the observed chaoticity does not necessarily have a physical meaning. 
In the Appendix we show, however, that LLE and ILE are indeed gauge 
invariant under time-independent 
gauge transformations in the temporal gauge.

Our goal is to clarify how the entropy grows during the time interval
when the gauge field configuration is still far from equilibrium 
but has already sampled a significant fraction of phase space.
Thus the KS entropy of interest should be defined as the sum of positive ILEs,
\begin{equation}
\frac{dS}{dt}=S_{KS}=\sum_{\lambda^\mathrm{ILE}_i>0} \lambda^\mathrm{ILE}_i
\ .
\end{equation}

\begin{widetext}
\begin{center}
\begin{table}[htb]
\caption{Parameter set for the modulated initial conditions. 
The column ``Parameter set'' gives the names of the parameter sets and 
the figures for which a parameter set was used.}
\begin{tabular}{llccccccc}
\hline
\hline
\multicolumn{2}{c}{Parameter set}
& lattice size &~~~~ $w$ (range of fluctuations) & $\epsilon_1$ & $\epsilon_2$ & $n$ & $m$ &$\epsilon$ (energy density) \\
\hline
M-F4 & (Fig.\ref{fig1}(a)) & $4^3$ &  $7.1\times10^{-2}$ 
	& 0  & 0 & - & - & $7.15\times10^{-3}$\\
M-B4 & (Fig.\ref{fig1}(b)) & $4^3$ &0 
	& $9.7\times10^{-2}$ & $10^{-4}$ &3 & 3 & $7.13\times10^{-3}$ \\
M-Bf4-1 & (Fig.\ref{fig1}(c) and \ref{fig4}) & $4^3$ & $10^{-4}$
	& $9.7\times10^{-2}$ & $10^{-4}$ &3 & 3 & $7.13\times10^{-3}$ \\
M-f4 & (Fig.\ref{fig1}(c$^\prime$))& $4^3$
	& $10^{-4}$ & 0 & 0 &- &-& $1.379\times 10^{-8}$ \\
M-Bf4-2	& (Fig.\ref{fig5}) & $4^3$ & $10^{-4}$
	& $6.859\times10^{-2}$ & $10^{-4}$ &3 & 3 & $3.529\times10^{-3}$ \\
M-Bf8 & (Fig.\ref{fig6}) & $8^3$ & $10^{-4}$ & $9.7\times10^{-2}$
	& $10^{-4}$ &3 & 3 & $3.528\times10^{-3}$ \\
\hline
\hline
\end{tabular}
\label{tab1}
\end{table}%


 \begin{table}[htb]
 \caption{Parameter set for constant $A$ initial conditions. }
 \begin{tabular}{llcccc}
 \hline
 \hline
 \multicolumn{2}{c}{Parameter set} & lattice size
	& $w$ (width of fluctuations) & magnetic field $B$  & $\epsilon$ (energy density)\\
 \hline
 C-Bf4 & (Fig.\ref{fig3} and \ref{fig11})& $4^3$
	& $5\times10^{-4}$ & $0.15$ & $1.45\times10^{-2}$\\
 C-Bf8 & (Fig.\ref{fig12})& $8^3$
	& $5\times10^{-4}$ & $0.15$ & $1.45\times10^{-2}$\\
\hline
\hline
\end{tabular}
\label{tab2}
\end{table}%
\end{center}
\end{widetext}

\subsection{Choice of initial conditions}

As mentioned in the Introduction,
realistic initial conditions for high-energy heavy-ion collisions are
based on the gauge field configuration created in the collision between two
color-glass condensates, in which both the chromomagnetic  and 
chromoelectric fields are parallel to the collision axis. 
These gauge field configurations serve as the starting point of the glasma.

For ``modulated initial conditions'' 
(denoted by ``M-'' in the names of the parameter set),
the spatial components of the gauge fields are given by 
an oscillating background field plus small fluctuations
\begin{align}
A^a_i (\vec r) =& \delta_{i 2}\left[\epsilon_1\sin\left(\frac{2x\pi}{N_x}\right)\right. \nonumber \\
& \left.+\epsilon_2\sin\left(\frac{2nx\pi}{N_x}\right)\sin\left(\frac{2mz\pi}{N_z}\right)\right] + \eta^a_i(\vec r)\, ,  \\
E^a_i(\vec r)&=0.
\label{modulated}
\end{align}
Here 
$\eta^a_i(\vec{r})$
is a random number 
that has indices for
color $a$
and spatial direction $i$ ($i=1,2,3$).

$\eta^a_i(\vec{r})$ is a uniformly distributed random number
ranging from $-w$ to $w$, 
where $w$ is the amplitude of the noise.  
The superscript $a$ denotes an adjoint color index ($a=1,\ldots,(N_c^2-1)$). 
In the present study, we study SU($N_c=2$). 
$N_x (N_z)$ is the lattice size in $x (z)$ direction. 
Recall that we use the temporal gauge and hence $A_0=0$.
The parameters for the modulated initial condition
are summarized in Table \ref{tab1}. 
We also include initial conditions with fluctuations
only
(i.e. without background field), namely M-F4 and M-f4, for comparison. 


When $\eta^a_i(\vec r)=0$, the initial magnetic fields are written as
\begin{align}
B_x^a&=\partial_y A_z^a - \partial_z A_y^a+g\epsilon^{abc}A_y^bA_z^c=-\partial_z A^a_y \\
&=-2m\frac{\pi}{N_z}\epsilon_2\sin\frac{2nx\pi}{N_x} \cos\frac{2mz\pi}{N_z}, \label{eqinitmf1}\\
B_y^a&=\partial_z A_x^a -\partial_x A_z^a +g\epsilon^{abc}A_y^bA_z^c=0, \\
B_z^a&=\partial_x A_y^a -\partial_y A_x^a+g\epsilon^{abc}A_y^bA_z^c=\partial_x A^a_y \\
&=\frac{2\pi}{N_x}\epsilon_1\cos\frac{2x\pi}{N_x}
 +\frac{2n\pi}{N_x}\epsilon_2\cos\frac{2nx\pi}{N_x} \sin\frac{2mz\pi}{N_z} .
\label{eqinitmf2}
\end{align}
Therefore, the background magnetic fields have $x$ and $z$ components,
and are modulated along the $x$ and $z$ direction.  

The other initial condition adopted here 
is the ``constant-$A$ initial condition'' 
(denoted by ``C-'' in the names of parameter sets), 
where 
constant gauge fields 
are used to produce constant magnetic fields (see~\cite{BSSS2011}), 
\begin{align}
\label{constantA}
A_i^a(\vec r)&=(\delta_{i2}\delta^{a3}+\delta_{i3}\delta^{a2})\sqrt{B/g}, \\
E_i^a(\vec r)&=0,
\end{align} 
where $g$ is the coupling constant. 
The vector potential gives the following magnetic field configuration in SU(2),
\begin{align}
B_z^1=-F_{yx}^1=g\epsilon^{1bc}A_y^b A_x^c=g(A_y^2 A_x^3-A_y^3A_x^2)=-B,
\end{align}
where the other components of the magnetic field are zero.
Since $g$ is an irrelevant parameter in CYM, we set $g=1$ throughout the paper. 
The parameters for the constant-$A$ 
initial condition are summarized in Table \ref{tab2}.
  
The constant-$A$ initial condition gives a constant chromomagnetic field
in one direction, and can be regarded as the simplest modeling 
of the initial glasma.
The $\epsilon_1$-dominant case of the modulated initial condition
($\epsilon_1 \gg \epsilon_2$)
also simulates the initial condition of the glasma 
with an additional amplitude oscillation in the $x$ direction:
The color-magnetic field aligns almost in the $z$ direction,
and a small longitudinal oscillation of the background field is introduced
by the small value of $\epsilon_2$. 
These features of the modulated initial conditions are 
advantageous for the investigation of
Nielsen-Olesen instabilities~\cite{Nielsen:1978rm}.
In a strong color-magnetic field,
the modulation in the color magnetic field direction
is most unstable \cite{FI2008,FII2009}.
 
While the color-electric field is absent in the present initial conditions,
the non-linear coupling between the chromo-magnetic fields and fluctuations
will generate chromo-electric fields very fast.
It would, of course, be desirable to incorporate chromo-electric 
fields in the initial condition
for a more comprehensive analysis of thermalization.
However, this is technically more involved
and beyond the scope of the present work, because 
one would have make sure that the initial condition satisfies 
Gauss' law for zero color charge.

\begin{figure*}[t]
\begin{center}
\begin{minipage}{1.1\linewidth}
\hspace{-2.7cm}
\includegraphics[width=0.45\textwidth]{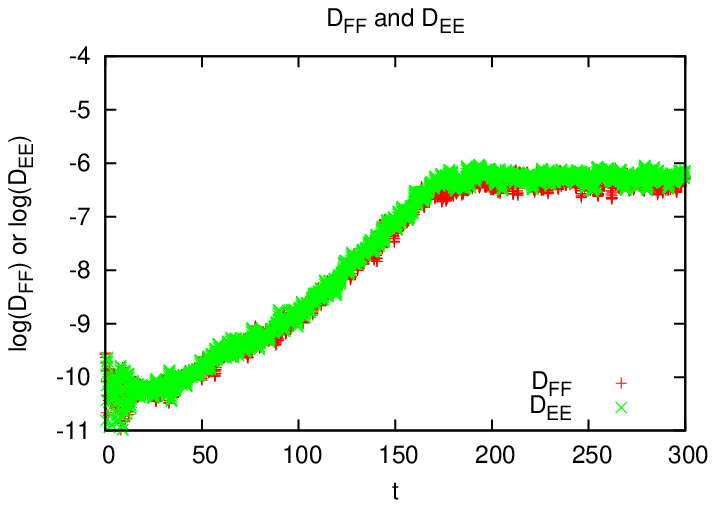}
\put(-140,55){{ \large \bf (a) only fluctuations}}
\put(-200,130){{ \Large \bf M-F4}}
\hspace{-0.5cm}
\includegraphics[width=0.45\textwidth]{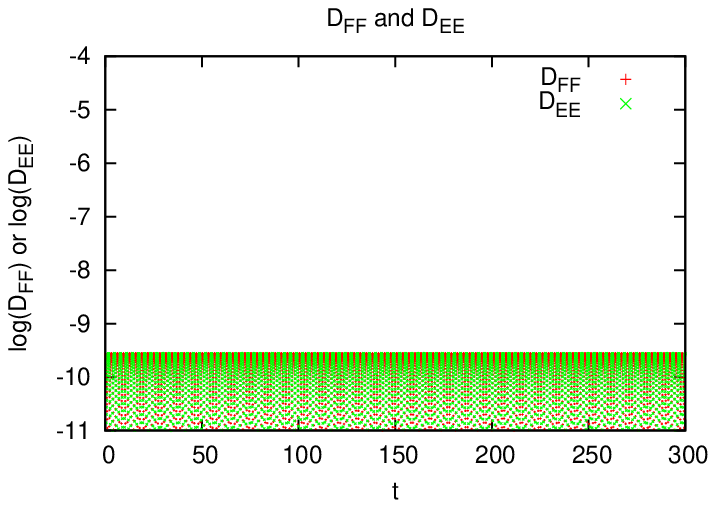}
\put(-180,80){{ \large \bf (b) without fluctuations}}
\put(-200,130){{ \Large \bf M-B4}}
\end{minipage}
\begin{minipage}{1.1\linewidth}
\hspace{-2.5cm}
\includegraphics[width=0.45\textwidth]{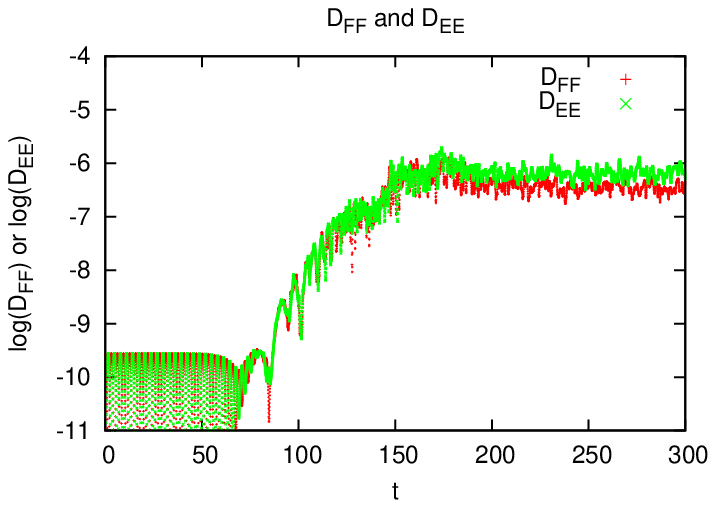}
\put(-120,50){{ \large \bf (c) background}}
\put(-135,35){{ \large \bf + tiny fluctuations}}
\put(-200,130){{ \Large \bf M-Bf4-1}}
\includegraphics[width=0.45\textwidth]{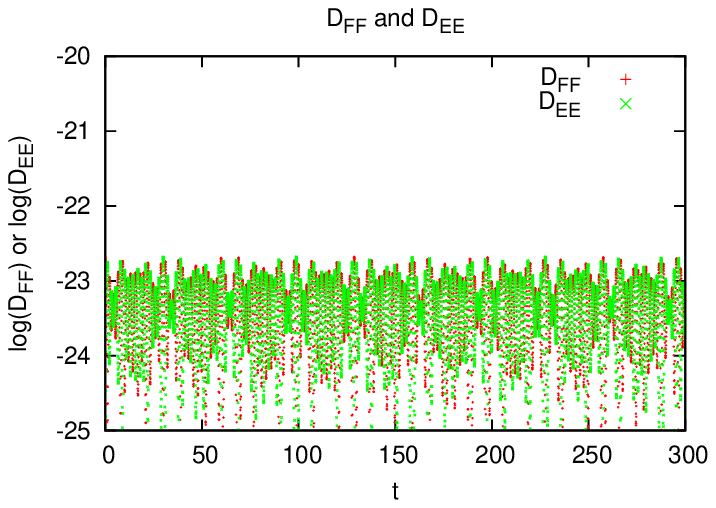}
\put(-185,100){{ \large \bf (c$^\prime$) only tiny fluctuations}}
\put(-200,130){{ \Large \bf M-f4}}
\end{minipage}
\caption{Time evolution of $D_{EE}$ and $D_{FF}$ with the initial conditions of 
(a) only fluctuations (parameter set M-F4 in Table \ref{tab1}),
(b) only background field (no fluctuations, M-B4),
(c) background plus tiny fluctuations (M-Bf4-1),
(c$^\prime$) only tiny fluctuations whose amplitude is the same
as that of (c) (M-f4).}
\label{fig1}
\end{center}
\end{figure*}

\begin{figure}[t]
\begin{center}
\includegraphics[width=0.45\textwidth]{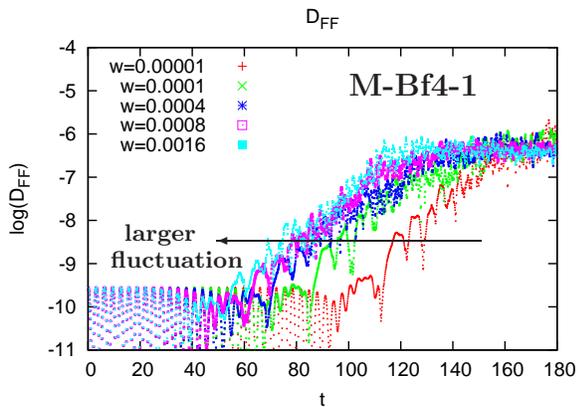}
\put(-90,120){{\large \bf M-Bf4-1}}
\put(-40,65){\vector(-1,0){100}}
\put(-175,65){\bf larger}
\put(-180,55){\bf fluctuation}
\caption{Dependence of $D_{EE}$ and $D_{FF}$ to the various size of amplitude of fluctuations, $w$, from modulated initial condition (M-Bf4-1).}
\label{fig2}
\end{center}
\end{figure}

\begin{figure}[t]
\begin{center}
\includegraphics[width=0.45\textwidth]{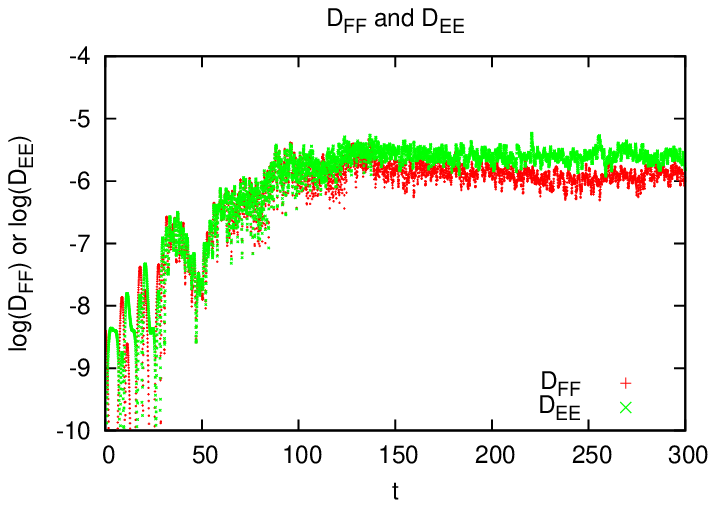}
\put(-90,60){{ \Large \bf C-Bf4}}
\caption{
Time evolution of $D_{EE}$ and $D_{FF}$ from
the initial condition
 C-Bf4 
where
the initial background color-magnetic field is constant (C-Bf4).
}
\label{fig3}
\end{center}
\end{figure}

\section{Chaoticity with glasma-like initial conditions}
\label{sec3}
\subsection{Time evolution of $D_{EE}$ and $D_{FF}$}

Here, we show the time evolution of 
the distances 
between two trajectories stemming from two adjacent points:
The parameter sets used for these analyses are summarized
in Tables \ref{tab1} and \ref{tab2}.
Figures \ref{fig1}(a), (b), (c) and (c$^\prime$) show 
$D_{EE}$ and $D_{FF}$ as functions of time for several different 
initial conditions,
M-F4, M-B4, M-Bf4-1 and M-f4 in Table~\ref{tab1}, respectively. 
The two (adjacent) starting points are related to each other by
\begin{align}
A^{\prime a}_i(t=0,\, \vec{r})&=1.005A^a_i(t=0,\, \vec{r})\ , \\
E^{\prime a}_i(t=0,\, \vec{r})&=1.005E^a_i(t=0,\, \vec{r})\ .
\end{align}
We note that a linear increase in the semi-logarithmic plots shows an 
exponential growth of the distance, which implies an exponential sensitivity 
of the trajectories to the initial value. This is a typical behavior showing 
the chaoticity of the system,
which leads to entropy production when combined with Husimi coarse-graining. 
The initial energy densities in the present cases 
are approximately the same except for the case shown
in Fig.~\ref{fig1}(c$^\prime$).

In Fig. \ref{fig1}(a), 
we show the time-evolution of the distances $D_{EE}$ and $D_{FF}$ with the initial condition M-F4, 
where the initial field is given by the fluctuation
with no background magnetic field. 
We can see the exponential growth of $D_{EE}$ and $D_{FF}$ from $t=30$,
which shows the chaoticity of the system.
After a time around $t=170$, these distances become saturated.   
We find almost the same saturation times for $D_{EE}$ and $D_{FF}$. 
This saturation property is understood to be due to the fixed total energy.

When the background magnetic fields are present in $z$-direction
but the fluctuation is absent in the initial condition (M-B4), 
the exponential growth of the distances does not manifest itself,
and they only
show a stationary oscillating behavior as shown in Fig.\ref{fig1}(b). 
In this case, thermalization is not expected to occur. 

Figure \ref{fig1}(c) shows the time-evolution of  
$D_{EE}$ and $D_{FF}$
with the background magnetic field plus very tiny fluctuations
in the initial condition (M-Bf4-1).
These $D$'s show an oscillatory behavior in the very early stage, 
then they start to grow exponentially at a certain time. 
In this setup, the onset time of the exponential growth is about $t=50$.

These results suggest that the solution of CYM
solely with glasma-like background field at initial time is unstable, but
the instability is triggered only when tiny but finite fluctuations are imposed
on top of the background field at initial time.
Figure \ref{fig1}(c$^\prime$) is the numerical result
solely with the initial fluctuations whose amplitude is the same as
that in Fig.~\ref{fig1}(c).
We can see that the distances do not show increasing behavior
at least until $t=300$.
Addition of such a tiny fluctuation is essential 
for the exponential growth
when it couples with the glasma-like background field
in the initial time, as seen in Fig.~\ref{fig1}(c).
This shows that the exponential growth of the distance between two
CYM solutions seen in Fig.~\ref{fig1}(c) is not caused by the fluctuations
themselves, but it is an inherent instability 
of the background field that is only triggered by the fluctuations.

We show the dependence of $D_{FF}$ on the  ratio of the fluctuations to the 
glasma-like background field in Fig.~\ref{fig2}:  
The larger the relative strength of the fluctuations, the earlier 
the onset time of the exponential growth and the saturation time of $D_{FF}$. 
This result is of phenomenological importance,
because we can predict the thermalization time once the ratio of
fluctuation to the background field is known.

Figure~\ref{fig3} shows $D_{FF}$ and $D_{EE}$
for a constant $A$ initial condition (C-Bf4).
We see that both the distances grow 
with more pronounced oscillations of a larger amplitude 
than  in the initial stage for the modulated initial condition,
but then again become saturated and almost constant on a 
time scale similar to that observed for the modulated initial 
condition (see Fig. 1(c)).

The oscillatory behavior in the initial stage can be understood by a linearized 
analysis of the CYM system, which shows that the leading-order time dependence
is given by a Jacobi elliptic function~\cite{BSSS2011}.

The increase of the distances for these different 
initial conditions indicates that chaotic behavior occurs 
irrespective of the details of the chosen initial conditions, 
as long as they have 
some (tiny) random fluctuations on some coherent background field.


\begin{figure*}[t]
\begin{center}
\begin{minipage}{1.0\linewidth}
\hspace{-2.7cm}
\includegraphics[width=0.40\textwidth]{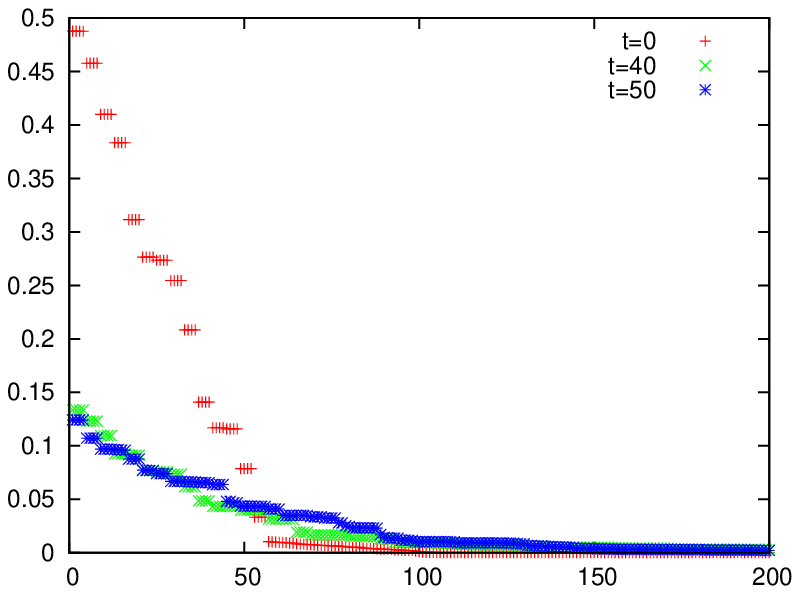}
\put(-150,100){{ \Large \bf M-Bf4-1}}
\put(-140,80){{ (energy density: 0.00713)}}
\hspace{-0.5cm}
\includegraphics[width=0.40\textwidth]{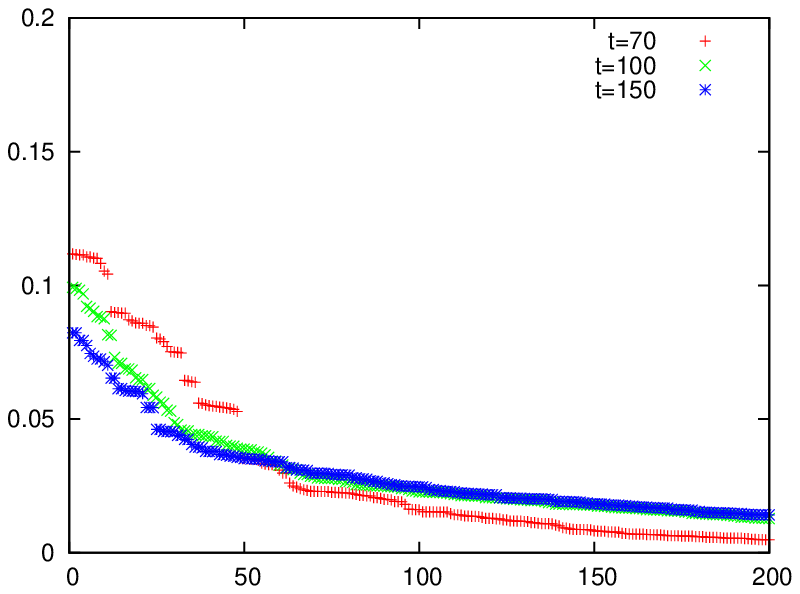}
\put(-150,100){{ \Large \bf M-Bf4-1}}
\put(-150,80){{  (energy density: 0.00713)}}
\end{minipage}
\caption{
 Time evolution of ILE for the ``modulated initial condition'' with M-Bf4-1 in Table \ref{tab1} ($V=4^3$).}
\label{fig4}
\end{center}
\end{figure*}

\begin{figure*}[t]
\begin{center}
\begin{minipage}{1.0\linewidth}
\hspace{-2.7cm}
\includegraphics[width=0.40\textwidth]{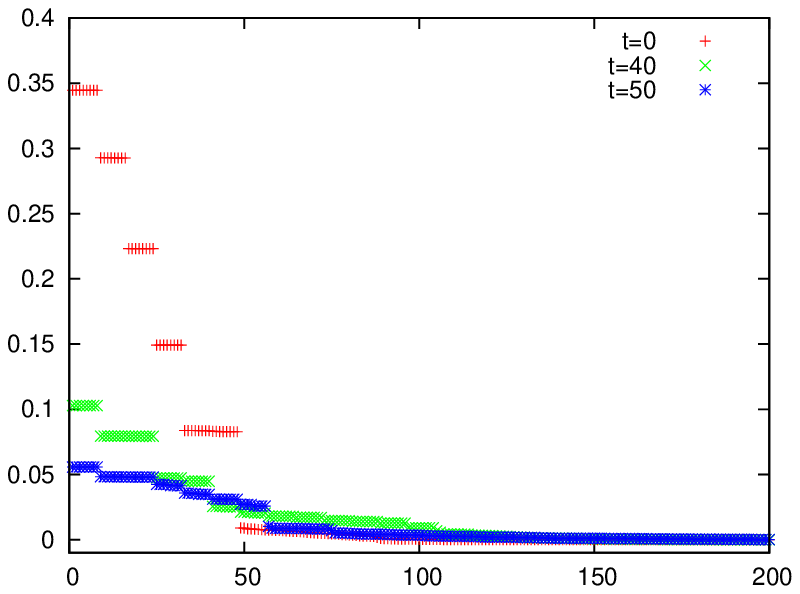}
\put(-150,100){{ \Large \bf M-Bf4-2}}
\put(-150,80){{ (energy density: 0.00353)}}
\hspace{-0.5cm}
\includegraphics[width=0.40\textwidth]{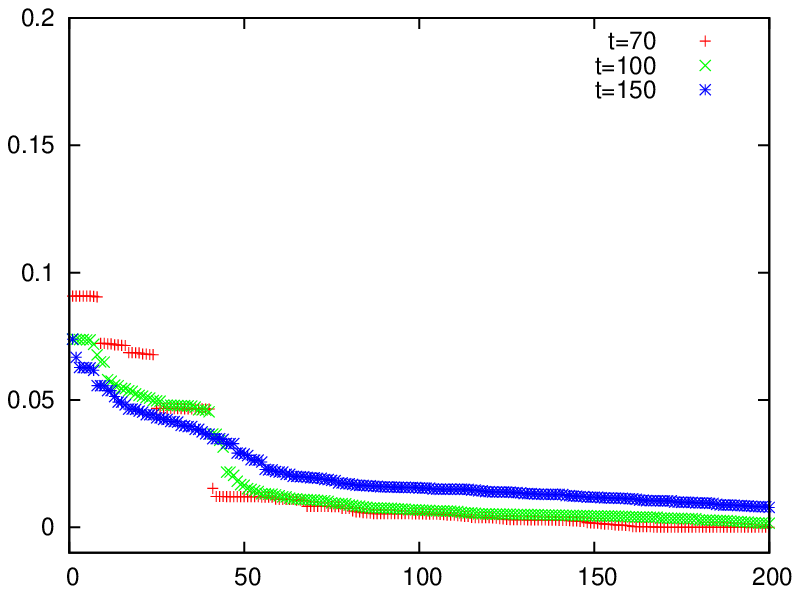}
\put(-150,100){{ \Large \bf M-Bf4-2}}
\put(-150,80){{(energy density: 0.00353)}}
\end{minipage}
\caption{
 Time evolution of ILE for the ``modulated initial condition'' with M-Bf4-2 in Table \ref{tab1} ($V=4^3$).}
\label{fig5}
\end{center}
\end{figure*}

\begin{figure*}[t]
\begin{center}
\begin{minipage}{1.0\linewidth}
\hspace{-2.7cm}
\includegraphics[width=0.40\textwidth]{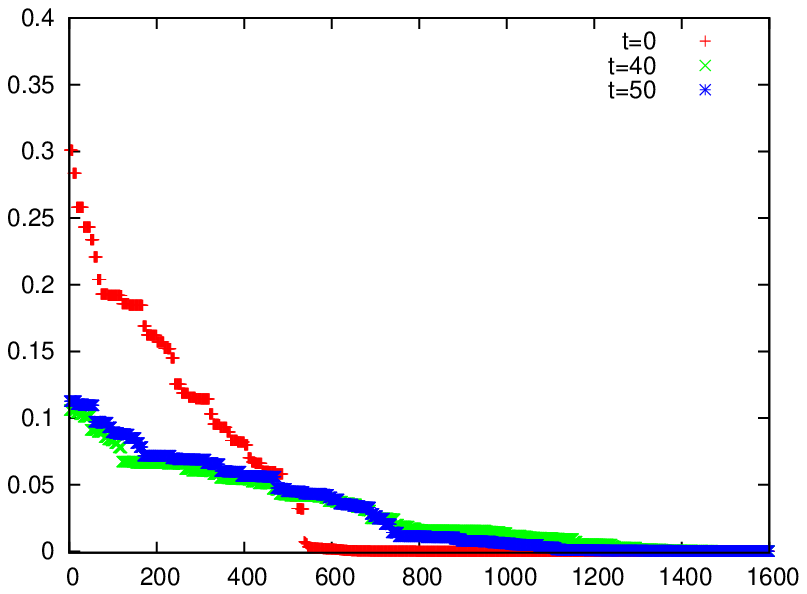}
\put(-150,100){{ \Large \bf M-Bf8}}
\put(-150,80){{(energy density: 0.00353)}}
\hspace{-0.5cm}
\includegraphics[width=0.40\textwidth]{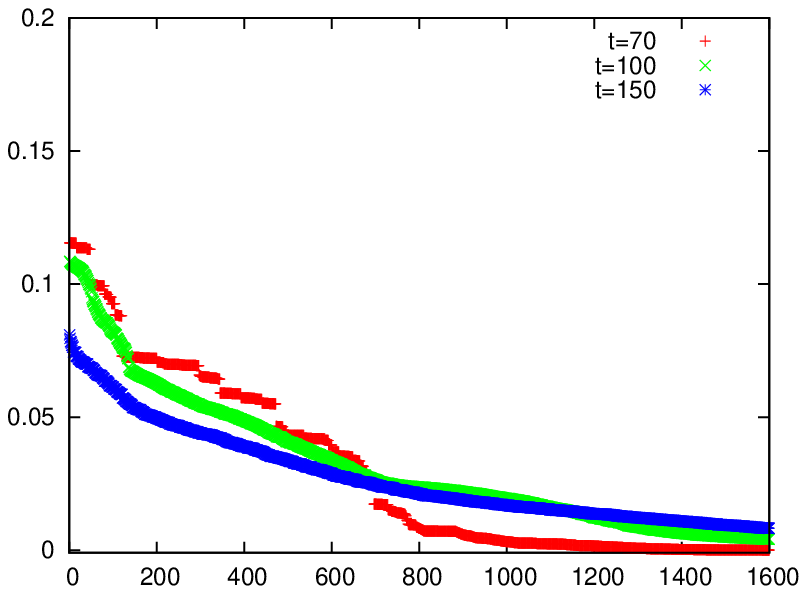}
\put(-150,100){{ \Large \bf M-Bf8}}
\put(-150,80){{(energy density: 0.00353)}}
\end{minipage}
\caption{
 Time evolution of ILE 
for the same initial condition as in Fig.~\ref{fig5} except for the lattice 
size ($8^3$ in this figure).  
The parameter set used in the calculation is M-Bf8 in Table \ref{tab1}.}
\label{fig6}
\end{center}
\end{figure*}

\begin{figure}[t]
\begin{center}
\includegraphics[width=0.40\textwidth]{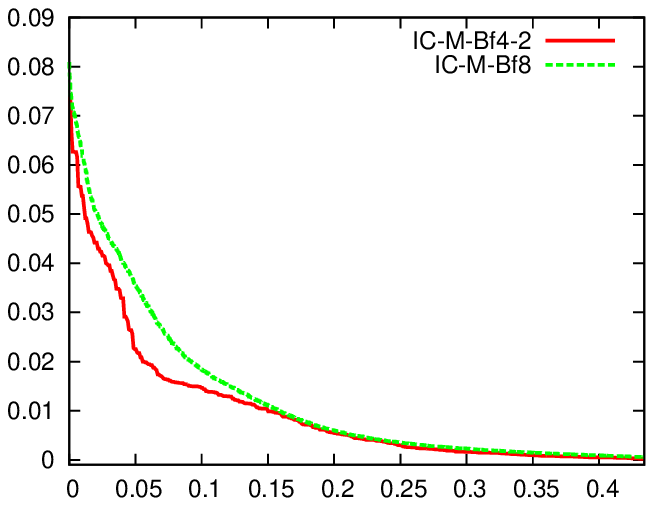}
\put(-120,90){{ \Large \bf $t=150$}}
\caption{
Dependence of ILE spectrum on lattice size.
The spectrum is plotted against the 
label of the ILEs divided by the total number of modes
in descending order of ILE. 
The red-solid line shows the spectrum 
on the $4^3$ lattice with the initial condition M-Bf4-2, 
and the green-dotted line shows that on the $8^3$ lattice
with the initial condition M-Bf8. 
These two initial conditions have the same energy density.}
\label{fig7}
\end{center}
\end{figure}

\subsection{Time evolution of the Lyapunov spectrum}

As mentioned before, the KS entropy is defined as the sum of all positive 
Lyapunov exponents and gives the entropy production rate.
Therefore, the spectrum of Lyapunov exponents in the late stage 
is of paramount importance for entropy production in a CYM system. 
We further expect that the sum of positive intermediate Lyapunov exponents
(ILEs) corresponds to the entropy production rate at a certain time.
If this is the case, the time evolution of the spectrum of ILEs describes
the time evolution of the thermalization process. 
Below we report our numerical results for the time evolution
of the spectrum of positive ILEs.

\begin{center}
{\bf Modulated initial condition}
\end{center}

Figure \ref{fig4} shows the time evolution of the spectrum of
ILEs for the parameter set M-Bf4-1 in Table \ref{tab1}. 
The vertical axis is the magnitude of the positive real ILEs
and the horizontal axis is the label of the ILEs in descending order.
Imaginary ILEs are plotted as zeros in these figures.   
The time slices for which the different spectra were obtained
are shown in the upper right-hand corner of the figures.

The figures show that the spectrum has a step-like shape at $t=0$, 
which implies the existence of degenerated, unstable, exponentially growing modes 
from the very beginning of the evolution. 
More importantly, the number of the positive Lyapunov exponents
is a finite fraction of the total number of degrees of freedom. 
This behavior actually could be expected, 
because the ILE's at $t=0$ coincide with the local Lyapunov exponents (LLEs)
defined as the eigenvalues of the Hessian, and
thus the existence of positive ILE's at $t=0$ should reflect that of 
the unstable modes revealed in linear response theory \cite{FI2008}.

The number of positive ILEs or unstable modes on the $4^3$ lattice is 
around 60 at $t=0$ and increases with time, while the magnitudes of the 
largest ILEs decrease during the time period of $0 < t \lesssim 100$. 
From a comparison with Fig.~\ref{fig1}(c), 
we find that the exponential growth of the $D$'s starts well after 
the spread of the spectrum of unstable modes. 
This observation suggests that a nonlinear analysis including mode-mode
coupling is necessary for the investigation of entropy production.
The spectrum at saturation contains a finite number of positive ILEs.
In fact, the number of positive ILE's now exceeds 200,
which constitutes a {\em macroscopic} number,
comparable to the total number of modes, $1152$. 
We conclude that the KS entropy is definitely nonzero and that 
entropy production is a sustained property of the glasma in CYM.

Figure \ref{fig5} shows the time evolution of the spectrum of ILEs 
for the parameter set M-Bf4-2, where the initial energy density is about half 
of that for M-Bf4-1. 
The feature of the spectrum is the same as that for M-Bf4-1: 
the step-like shape at the initial time and 
the time dependence of the ILE spectrum. 
A difference is that the magnitude is 
smaller than for M-Bf4-1 at any point along the horizontal axis. 
This fact shows that KS entropy is larger for a larger energy density. 
We will show the value of the KS entropy in Table \ref{tab3}.

Figure \ref{fig6} shows the evolution
of the ILE spectrum 
in a larger ($8^3$) volume, 
with the parameter set M-Bf8 shown in Table \ref{tab1}.
The plot style is the same as in Fig.~\ref{fig4}.
The spectrum 
again starts from a somewhat discontinuous 
and step-like shape at $t=0$ and then becomes a smooth shape 
quite similar to that found on the smaller ($4^3$) lattice.
The transition between the two regimes occurs at $t \simeq 50$.
It should be noted that the total number of degrees of freedom,
$18N^3$ for a $N^3$ lattice, is quite different for the two lattices.
Again, the Lyapunov spectrum becomes stable after $t\simeq 100$,
containing a large number of positive ILE's, approximately 4500, which
constitutes a macroscopic fraction of the total number of degrees of
freedom, which is 9216.
The similarity with the results for the smaller lattice extends to 
quantitative details. For example, the value of the largest ILE is near 0.08
for both lattices, and the values of the ILEs for the $17 \%$ most unstable
modes exceed 0.01 in both cases.

In Fig.\ref{fig7}, the ILE spectra for M-Bf4-2 and M-Bf8 are compared 
at late time, $t=150$, where the spectrum 
is approximately stable. 
We can see that the magnitude of the ILEs for the $8^3$ lattice
is always larger than for $4^3$ at any point along the horizontal axis. 
In addition, the spectra 
coincide for large values on the horizontal axis.
Based on this analysis of volume-dependence, we expect
that the ILE spectrum 
stays finite in infinite volume.

In Table~\ref{tab3}, the time dependence of the KS entropy 
divided by volume, 
$s_{\rm KS}\equiv S_{\rm KS}/V=\sum_{\lambda_i>0} \lambda_i/V$,
is shown for each initial condition, M-Bf4-1, M-Bf4-2, and M-Bf8.
Note that M-Bf4-2 and M-Bf8 have the same initial energy density.
For the initial condition M-Bf4-1, the value of
$s_{\rm KS}\equiv S_{\rm KS}/V$ is about 0.23 at the initial time, 
drops slowly with time and  finally  settles
around 0.12 at $t\simeq 100$.
Thus, the KS entropy is surely positive definite even at late times,
implying that the entropy is produced. 
A similar behavior is obtained for the initial condition M-Bf4-2
which has an energy density about half of that of M-Bf4-1;
$s_{\rm KS}$ starts from a value around 0.15,
and the stable final value around 0.08 is reached at $t\simeq 150$.
The larger the energy density, the larger the value of the KS entropy.
In the previous study, Ref.~\cite{KMOSTY2010}, 
the authors found that $s_{\rm KS}$ scales as the fourth root of 
energy density, $\epsilon^{1/4}$, 
for random initial magnetic fields. 
For the initial conditions M-Bf4-1 and M-Bf4-2, 
the ratio of KS entropy, $0.122/0.079\simeq 1.54$, is not so close to 
the fourth root of the ratio of the energy density 
$\epsilon$, $(0.00706/0.00353)^{1/4}\simeq 1.19$. 
As the setting in this case is not isotropic, 
it is not clear that all dimensioned parameters 
should simply scale with powers of $\epsilon^{1/4}$. 
It could also be that the volume studied here ($4^3$) 
is insufficient for a reliable scaling analysis.

For the larger volume, M-Bf8, $s_{\rm KS}$ is about 0.14
at initial time and about 0.10 at $t=150$. 
$s_{\rm KS}$ surely survives in a larger volume,
and entropy production in the infinite-volume limit can be expected.

\begin{center}
{\bf Constant-$A$ initial condition}
\end{center}

\begin{figure*}[t]
\begin{center}
\begin{minipage}{1.0\linewidth}
\hspace{-2.7cm}
\includegraphics[width=0.40\textwidth]{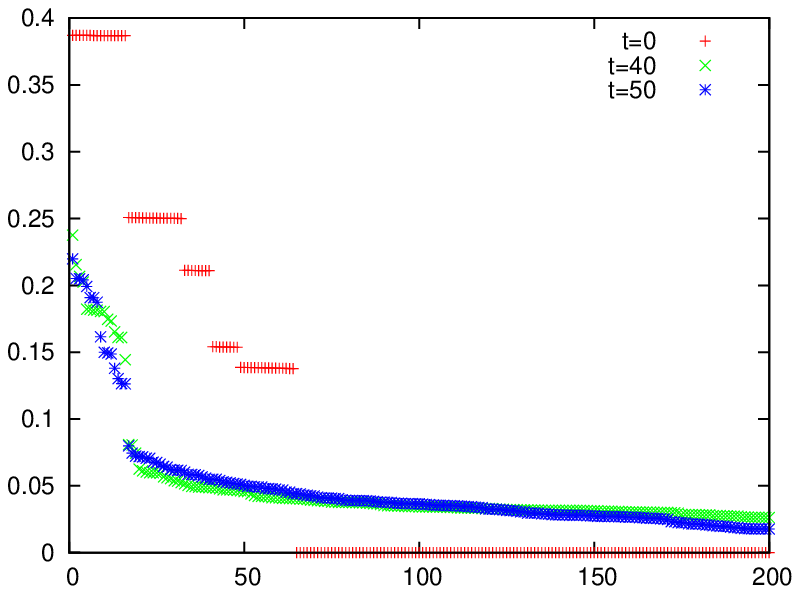}
\put(-140,110){{ \Large \bf C-Bf4}}
\hspace{-0.5cm}
\includegraphics[width=0.40\textwidth]{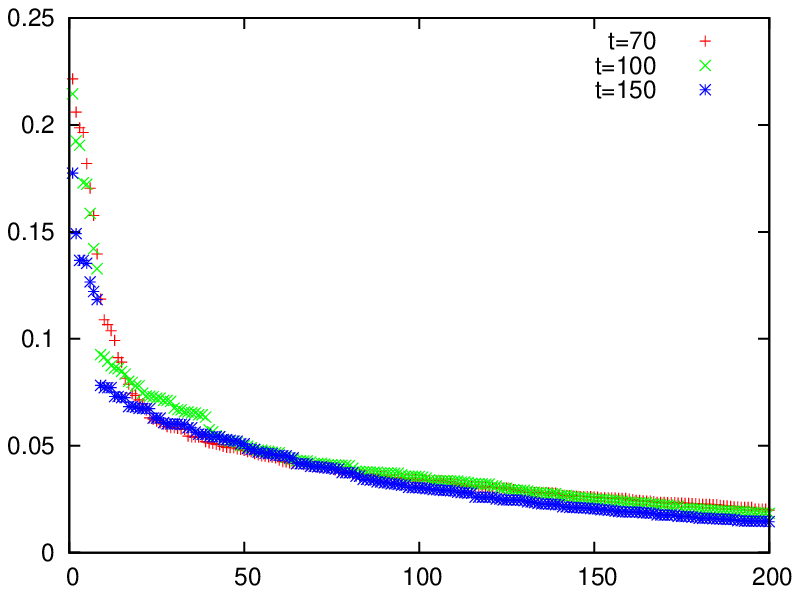}
\put(-140,110){{ \Large \bf C-Bf4}}
\end{minipage}
\caption{Time evolution of the intermediate Lyapunov exponent spectrum 
for the ``constant $A$ initial condition'' on the $V=4^3$ lattice.
The parameter set used in the calculation is C-Bf4 from Table \ref{tab2}.}
\label{fig11}
\end{center}
\end{figure*}

\begin{figure*}[t]
\begin{center}
\begin{minipage}{1.0\linewidth}
\hspace{-2.7cm}
\includegraphics[width=0.40\textwidth]{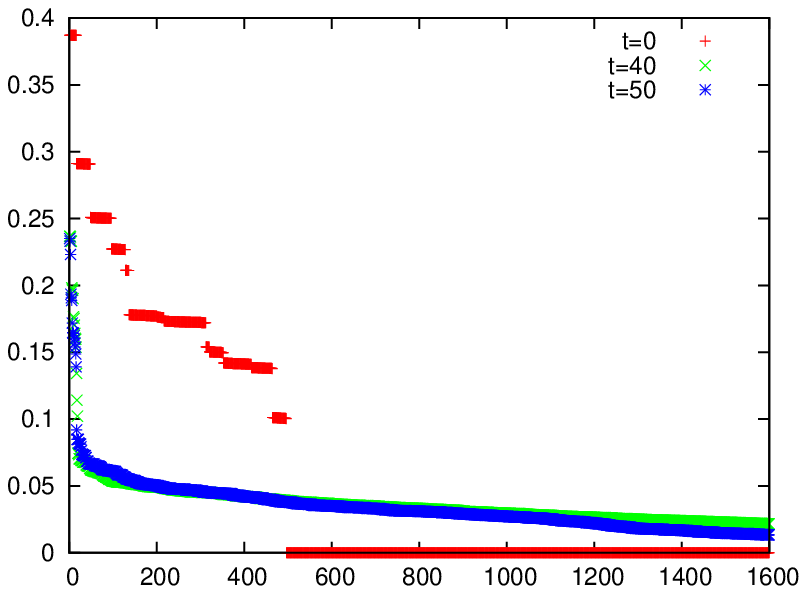}
\put(-140,110){{ \Large \bf C-Bf8}}
\hspace{-0.5cm}
\includegraphics[width=0.40\textwidth]{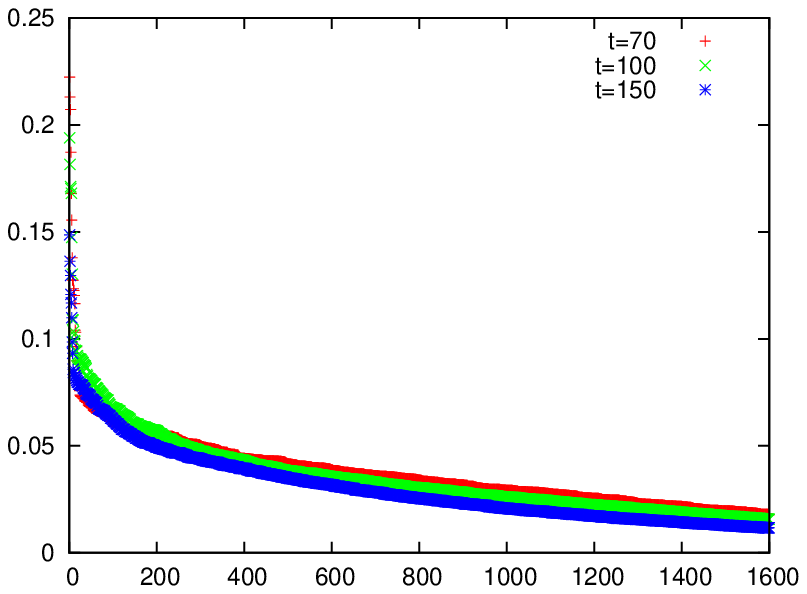}
\put(-140,110){{ \Large \bf C-Bf8}}
\end{minipage}
\caption{Time evolution of intermediate Lyapunov exponent spectrum 
for the ``constant $A$ initial condition'' on the  $V=8^3$ lattice.
The parameter set used in the calculation is C-Bf8 from Table \ref{tab2}.}
\label{fig12}
\end{center}
\end{figure*}

We show the time evolution of the ILE spectra
obtained from the constant-$A$ initial conditions
with tiny fluctuation on $4^3$ and $8^3$ lattices
in Figs.~\ref{fig11} and \ref{fig12}, respectively. 
At $t=0$, the spectra have the step-like structure
similar to that from the modulated initial condition,
except for the detailed structures.
At later time, this discontinuity of the structure disappears and the spectrum 
becomes smooth. After about $t=100$, the spectrum 
becomes stable,
and again the fraction of positive ILEs is finite.
The behavior on the $8^3$ lattice 
is quantitatively quite similar to that on the $4^3$ lattice,
when the horizontal axis is rescaled by the total number of modes.

Comparison of the two cases we have studied,
modulated and constant-$A$ initial conditions,
which are chosen to mimic the glasma initial condition,
reveals that the ILE spectra at late times are quite similar: 
Except for a few modes that remain very unstable ($\lambda > 0.1$)
in the case of the constant-$A$ initial condition, the largest ILE's take 
values around 0.08, and the modes at the $17^\mathrm{th}$ percentile 
of all modes (the $200^\mathrm{th}$ and $1600^\mathrm{th}$ modes for the 
$4^3$ and $8^3$ lattices, respectively) have ILEs around 0.01. 
In both cases also a much larger fraction of all modes becomes 
unstable at late times than in the initial phase of the evolution.
The agreement suggests that entropy production occurs in the CYM 
dynamics for any glasma-like initial condition and that a large part of the 
entropy production is caused by the chaoticity of the gauge field dynamics 
manifested after some time, 
rather than by the instability of the specific gauge field configuration
in the initial stage.


\begin{center}
\begin{table}
\begin{tabular}{c|c|c|c}
 \hline
 \hline
\multicolumn{4}{c}{$s_{\rm KS}\equiv S_{\rm KS}/V$}\\
\hline 
  & M-Bf4-1 & M-Bf4-2 & M-Bf8 \\
  & {\scriptsize $\epsilon=7.13\times10^{-3}$}
  & {\scriptsize $\epsilon=3.53\times10^{-3}$}
  & {\scriptsize $\epsilon=3.53\times10^{-3}$}     
\\
 \hline
 $t$=0    & 0.226&  0.151 & 0.140 \\ 
 40  & 0.088 &  0.063 & 0.095 \\ 
 50  &0.094 &  0.042 & 0.096  \\ 
 70  & 0.098&  0.052 & 0.092  \\ 
 100 & 0.119&  0.054 & 0.106  \\ 
 150 & 0.122 &  0.079 & 0.099 \\ 
\hline
\hline
\end{tabular}
\caption{Time dependence of Kolmogorov-Sina\"i entropy divided by volume
for initial conditions, M-Bf4-1, M-Bf4-2 and M-Bf8.
$\epsilon$ denotes the energy density.
\label{tab3}}
\end{table}%
\end{center}

\section{Summary and Concluding Remarks}
\label{sec5}

We have studied the thermalization process of classical Yang-Mills fields 
starting from semi-realistic glasma-like initial conditions characteristic
for relativistic heavy-ion collisions.
We focused on the chaotic behavior and entropy production
of the classical field theory. 
For this purpose we determined all Lyapunov exponents for the
theory discretized on a cubic spatial point lattice. The sum of 
all positive Lyapunov exponents, i.e. the Kolmogorov-Sina\"i entropy,
gives the rate of entropy production. 

We considered two types of glasma-like initial conditions.
One is the ``modulated initial condition'', where initial chromomagnetic fields 
$B^a_i$ are spatially modulated along the $z$ and $x$ axes. 
In the absence of fluctuations, this field configuration is color-independent. 
The other initial condition is the ``constant-$A$ initial condition'',  where the gauge potentials 
$A^a_i$ and chromo-magnetic fields $B^a_i$ are both constant, but color-dependent,
and only the $z$-component of the chromomagnetic field in one color
direction is nonzero.
For both types of initial conditions we added small random fluctuations
to describe the noise present in the glasma fields due 
to the quantum fluctuations in the two colliding nuclei. 

We have found that
when the gauge field is given by the modulated initial condition 
without any fluctuations, the distance between two trajectories starting from 
adjacent points shows oscillatory behavior and exponential growth is absent.
When small fluctuations are present on top of the background field,
the initial oscillatory behavior of the distance terminates after a short time 
and the exponential growth of the distance becomes robustly visible. 
The onset time of exponential growth depends on the ratio of
the amplitude of the fluctuation to the strength of the chromomagnetic field.
The fact that we do not see any exponential growth in the absence of 
initial fluctuations demonstrates that our numeric treatment introduces 
no fluctuations of relevant size. 

For the modulated initial condition with small fluctuations 
positive Lyapunov exponents are present from the
beginning. This is a reflection of the unstable modes revealed in the
linear response analysis of glasma fields. We found that the number of 
modes with positive Lyapunov exponents increases substantially during 
time evolution, and that the Lyapunov spectrum reaches a stable shape
when a macroscopic fraction of the modes has positive Lyapunov exponents. 
This implies that a bulk number of field modes contributes to 
the KS entropy
and that the entropy production rate per unit volume is non-zero in CYM.

The dependence of the Lyapunov spectrum on the size of the spatial volume 
is found to be weak if the mode number is rescaled by the total 
number of modes, which suggests that extensive entropy production occurs
in the infinite volume limit in CYM as first found by Bolte {\em et al.}
\cite{Bolte:1999th}.

The chaotic behavior of CYM with the noisy ``constant-$A$ initial condition'' 
is similar to that observed for the modulated initial condition with some
modifications during the initial phase of the time evolution.
After an initial increase of the distance accompanied with some oscillatory 
behavior, the distance shows an exponential growth, and the time evolution 
of the spectrum of ILEs shows a qualitatively similar behavior as that 
seen for the modulated initial condition. 
The number of positive Lyapunov exponents is again found to be large, 
corresponding to many unstable {\em bulk} modes,
and the entropy production rate grows approximately linear with volume 
also for the constant $A$ initial condition. 

The important conclusion to be drawn from our present analysis is that
entropy production occurs in a robust way 
{\it at the classical field level} 
starting from semi-realistic initial conditions. 
While the classical field description of the glasma becomes invalid when
thermal equilibrium is approached, the pre-equilibrium dynamics of the 
glasma can be simulated by classical gauge field equations, 
which allows to estimate the thermalization time. 
We therefore focused our study on the intermediate time regime, where
the ILEs characterize the dynamical instability of the gauge field. 
Our study suggests that the magnitude of the delay time 
before the start of linear entropy growth 
depends substantially on the chosen initial condition while the
entropy growth rate itself is affected at most mildly. Substantially
more systematic studies are needed to decide whether there exist realistic 
scenarios for which the thermalization time is as small as
1 fm/c, which is the value advocated by the comparison of 
hydrodynamical simulations with heavy ion data.

As the gauge field approaches equilibrium and quantum effects become
important by providing a physical cut-off to the ultraviolet divergences 
of the classical thermal field theory, the evolution of the glasma can 
be described by viscous hydrodynamics.
It is an interesting question whether the two effective theories of the
dynamics of an equilibrating quark-gluon plasma can be directly
joined or whether there is a need for an additional
formalism interpolating between these two regimes. 

\section*{ACKNOWLEDGMENTS}
The calculations were performed mainly by using the NEC-SX9 at
Osaka University. 
This work was supported in part by
 Grant-in-Aid for Scientific Research from
 the Japan Society for the Promotion of Science (JSPS)
 and the Ministry of Education, Culture, Sports, Science and Technology
 of Japan (MEXT)
 (Nos.
 20540265, 
 Innovative Areas (No. 2004: 23105713, and No. 2404: 24105001, 24105008), 
 23340067, 
 24340054, 
 24540271
),
 by the Yukawa International Program for Quark-Hadron Sciences,
 by a Grant-in-Aid for the global COE program
``The Next Generation of Physics, Spun from Universality and Emergence''
from MEXT and by BMBF (06RY7195).

\appendix
\def\thesection{}
\section{Proof of the gauge invariance of Lyapunov spectra}
We here show that the local/intermediate Lyapunov exponents are gauge invariant under the remaining gauge transformations in temporal gauge, i.e., time-independent gauge transformations. 
The Hessian $\Hess(t)$ in a concrete expression is
\begin{align}
\Hess(t)=
\begin{pmatrix}
{\bf 0} & {\bf 1} \\
-H_{AA}(t) & {\bf 0} \\
\end{pmatrix}
,
\end{align}
where 
$(H_{AA}(t))_{iax,jby}\equiv{\delta^2 H}/{\delta A_i^a(\vec x,t)}{\delta A_j^b(\vec y,t)}$
is a second derivative in terms of gauge fields. 
Gauge fields $A_i^a$ are transformed according to
$A'{_i^a}(\vec x,t) = (\Omega(\vec x) A_i(\vec x,t))^a+W_i^a(\vec x,t)$,
with a time-independent orthogonal matrix $\Omega(\vec x)$, i.e., $\Omega(\vec x)\Omega^T(\vec x)={\bf 1}$.
$W_i^a(\vec x,t)$ is a possible term independent of gauge fields.

Paying attention to the chain rule,
\[
\frac{\delta}{\delta A'{_i^a}(\vec x,t)}
=
\frac{\delta}{\delta A{_i^b}(\vec x,t)}\frac{\delta A{_i^b}(\vec x,t)}{\delta A'{_i^a}(\vec x,t)}
=
\frac{\delta}{\delta A{_i^b}(\vec x,t)}\left(\Omega^T(\vec x)\right)_{ba}
\]
the transformation property of $H_{AA}$ for a  time-independent gauge 
transformation $\Omega(\vec x)$ in the adjoint representation of SU(N) is found to be
\begin{eqnarray}
H_{AA}(t)&=&{\delta^2 H}/{\delta A_i^a(\vec x,t)}{\delta A_j^b(\vec y,t)}\nonumber \\
\rightarrow
H'_{AA}(t)&=&
{\delta H^2}/{\delta A'{_i^{a}}(\vec x,t)}{\delta A'{_j^b}(\vec y,t)} \nonumber \\
&=&(\Omega H_{AA}(t) \Omega^T)_{iax,jby}.
\end{eqnarray}
(Note that the Hamiltonian $H$ is gauge invariant.)
Defining an orthogonal martix
\begin{equation}
\overline{\Omega}(\vec x)\equiv
\begin{pmatrix}
\Omega(\vec x) & {\bf 0} \\
{\bf 0} & \Omega(\vec x)
\end{pmatrix},
\end{equation}
the gauge transformation of ${\cal H}(t)$ can be expressed as
\begin{equation}
{\cal H}(t) \rightarrow {\cal H}'(t)=\overline{\Omega}{\cal H}(t)\overline{\Omega}^T
\end{equation}
in a similar manner.
Taking orthogonality and time-independence of $\overline{\Omega}(\vec x)$
into account,
the local and intermediate Lyapunov spectra of the Hessian ${\cal H}(t)$
are easily shown to be gauge invariant.
In fact, a time-ordered product of the Hessian
\begin{align}
U(t,t+\Delta t)
=\,& {\cal T} \left[\exp\left( \int_t^{t+\Delta t} \Hess(t+t') dt' \right)	\right]
\end{align}
transforms as
\begin{eqnarray}
U(t,t+\Delta t)
&\rightarrow&
U'(t,t+\Delta t) \nonumber \\
&=&
{\cal T} \left[\exp\left( \int_t^{t+\Delta t} \overline{\Omega}\Hess(t+t')\overline{\Omega}^T dt' \right)	\right] \nonumber \\
&=&
\overline{\Omega}{\cal T} \left[\exp\left( \int_t^{t+\Delta t} \Hess(t+t') dt' \right)	\right]\overline{\Omega}^T \nonumber \\
&=&
\overline{\Omega} U(t,t+\Delta t)  \overline{\Omega}^T,
\end{eqnarray}
and gives the same (gauge invariant) intermediate Lyapunov spectra.


\end{document}